\RequirePackage{amsthm}

\documentclass[sn-mathphys-num]{sn-jnl}

\usepackage{graphicx}%
\usepackage{multirow}%
\usepackage{amsmath,amssymb,amsfonts}%
\usepackage{amsthm}%
\usepackage{mathrsfs}%
\usepackage[title]{appendix}%
\usepackage{xcolor}%
\usepackage{textcomp}%
\usepackage{manyfoot}%
\usepackage{booktabs}%
\usepackage{algorithm}%
\usepackage{algorithmicx}%
\usepackage{algpseudocode}%
\usepackage{listings}%
\usepackage{lmodern}

\theoremstyle{thmstyleone}%
\theoremstyle{thmstyletwo}%

\theoremstyle{thmstylethree}%

\begin{document}

\title[Sentiment Analysis of ML Projects: Bridging Emotional Intelligence and Code Quality]{Sentiment Analysis of ML Projects: Bridging Emotional Intelligence and Code Quality} 

\author{\fnm{Md Shoaib} \sur{Ahmed}}\email{mdshoaibahmed@u.boisestate.edu}
\equalcont{These authors contributed equally to this work.}
\author{\fnm{Dongyoung} \sur{Park}}\email{youngpark@u.boisestate.edu}
\equalcont{These authors contributed equally to this work.}
\author*{\fnm{Nasir U.} \sur{Eisty}}\email{nasireisty@boisestate.edu}

\affil{\orgdiv{Department of Computer Science}, \orgname{Boise State University}, \orgaddress{\state{Idaho}, \country{USA}}}



\abstract{This study explores the intricate relationship between sentiment analysis (SA) and code quality within machine learning (ML) projects, illustrating how the emotional dynamics of developers affect the technical and functional attributes of software projects. Recognizing the vital role of developer sentiments, this research employs advanced sentiment analysis techniques to scrutinize affective states from textual interactions such as code comments, commit messages, and issue discussions within high-profile ML projects. By integrating a comprehensive dataset of popular ML repositories, this analysis applies a blend of rule-based, machine learning, and hybrid sentiment analysis methodologies to systematically quantify sentiment scores. The emotional valence expressed by developers is then correlated with a spectrum of code quality indicators including the prevalence of bugs, vulnerabilities, security hotspots, code smells, and duplication instances. Findings from this study distinctly illustrate that positive sentiments among developers are strongly associated with superior code quality metrics—manifested through reduced bugs and lower incidence of code smells. This relationship underscores the importance of fostering positive emotional environments to enhance productivity and code craftsmanship. Conversely, the analysis reveals that negative sentiments correlate with an uptick in code issues, particularly increased duplication and heightened security risks, pointing to the detrimental effects of adverse emotional conditions on project health.}




\maketitle

\section{Introduction}
Understanding the complexities of human communication depends on sentiment - the subtle expression of human emotions and viewpoints. Sentiments in text data can vary widely, from vibrant expressions of positivity to down tones of negativity, which are often influenced by context, tone, and cultural traditions. The growing field of sentiment analysis (SA) combines linguistics and artificial intelligence to uncover the emotional degree in written content~\cite{taboada2016sentiment, ahmed2021detecting}. Its goal is to decode the emotional subtext embedded within the text. SA has many applications in different fields, such as business, social media, education, and health. It assists us in gaining insights into people's thoughts and feelings regarding different topics, services, products, and events. Furthermore, SA can play a significant role in enhancing the way we communicate and interact with others.

Software engineering (SE) is a human-intensive activity involving technical skills and emotional and social aspects. Developers often express their sentiments and opinions in various textual sources related to SE, such as code comments, commit messages, issue reports, pull requests, documentation, and online forums~\cite{lin2018sentiment}. Sentiment analysis is a technique that can automatically identify and extract emotions and opinions from natural language texts. By applying sentiment analysis to software engineering texts, we can gain insights into the affective states, attitudes, and preferences of developers and stakeholders, as well as the code's readability, maintainability, and usability. Sentiment analysis can also help us improve code quality and developers' emotional intelligence in SE projects. 

Machine learning (ML) initiatives are considered the foundation of innovation in the current AI era, contributing to progress across diverse industries and transforming how we engage with technology. ML has become a key driver of innovation and transformation in various domains, such as business, health, education, and entertainment. ML projects focus on creating and managing ML models that can address intricate challenges and benefit users and stakeholders. However, ML projects face many challenges, such as data quality, scalability, interpretability, and ethics. The significance of sentiment and emotions is often neglected in ML projects. By applying sentiment analysis to ML projects, we can gain insights into the affective states, attitudes, and preferences of developers, customers, and users and the readability, maintainability, and usability of ML models. Sentiment analysis can also help us improve the quality of ML models and the emotional intelligence of developers in ML projects by providing feedback, guidance, and motivation.

There are various methods to perform sentiment analysis on text, such as rule-based, machine learning, and hybrid methods. The rule-based method uses a set of predefined rules and dictionaries to assign sentiment scores to words and phrases in the text. The comprehensive sentiment of the text is measured by combining the scores of the individual components~\cite{asghar2017lexicon}. Machine learning approaches utilize supervised or unsupervised learning algorithms to train a sentiment classifier on a large corpus of labeled or unlabeled text data. The classifier can then predict the sentiment of new text based on the learned features and patterns~\cite{hasan2018machine}. Finally, hybrid methods combine rule-based and machine-learning methods to leverage the strengths of both approaches. For example, a hybrid method can use a rule-based method to generate sentiment labels for unlabeled data and then use a machine learning method to refine and improve the labels~\cite{le2019hybrid}.

Code quality gauges the worth of a particular codebase, program, or software. Generally, code is considered to be of high quality if it is straightforward to read and well-documented by the developer. High-quality code usually adheres to these criteria: it's functional, uniform, simple to comprehend, satisfies client requirements, can be tested, is reusable, devoid of bugs and errors, secure, and thoroughly documented. Ensuring the quality of the codebase is crucial as it precisely affects the comprehensive quality of the software \cite{keuning2023systematic}. The degree of safety, security, and reliability of the codebase is determined by the quality of the code, making it an essential aspect for many development teams, particularly those working on safety-critical systems.

The impact of sentiments on code quality represents a novel and fascinating frontier in software engineering. Sentiment analysis can also be utilized to assess and enhance software code quality in software engineering. Developers often express their sentiments and opinions in various textual sources related to software engineering, such as code comments, commit messages, issue reports, pull requests, documentation, and online forums. By applying sentiment analysis to these sources, we can gain insights into the affective states, attitudes, and preferences of developers and stakeholders, as well as the code's readability, maintainability, and usability. Sentiment analysis can also help us improve the quality of code and the emotional intelligence of developers in software engineering projects, especially in ML projects, where the complexity and uncertainty of the tasks require high levels of creativity, collaboration, and communication.
\section{Related Work}

\textbf{Sentiment analysis in Software Engineering.} 
The relationship between emotions and software engineering is intricate, highlighting the complex and collaborative nature of the field. Software development, a multifaceted process, involves various stakeholders, each contributing their distinct insights. Research in this area has focused on understanding and leveraging these emotional dynamics to improve team interactions and project outcomes. Biswas et al.~\cite{biswas2020achieving} have made significant strides by implementing BERT for sentiment analysis in software engineering texts. Their findings demonstrate enhanced accuracy in sentiment detection within software-related documents, facilitating better tool development for API recommendations and improved communication among developers. Exploring deeper into the emotional undercurrents within software teams, Gachechiladze et al.~\cite{Gachechiladze} investigate the specific expressions of anger among team members. Their research highlights the critical need to distinguish between anger directed at oneself, others, and objects. Identifying the direction of anger can significantly enhance collaborative tools, helping teams address internal challenges, manage community dynamics, and prioritize actions based on emotional signals. 

In a novel approach, Guzmán and Brügge~\cite{Guzman} propose the use of quantitative emotion summaries to foster emotional awareness within software development teams. By analyzing and summarizing emotions from collaborative artifacts, their study reveals a strong correlation between these summaries and the overall emotional climate of a project, paving the way for new methods to enhance emotional sensitivity in team settings. Cao and Park~\cite{Cao2017UnderstandingGE} reavels the emotional dynamics of agile teams, examining how emotions triggered by project and individual goals can influence behavior and project outcomes. Their work underscores the impact of emotional responses on the agile process and suggests ways to harness these insights for better team performance. Additionally, Fountaine and Sharif~\cite{Fountaine} discuss the integration of emotional intelligence practices within software development environments. They argue that understanding and managing emotions within teams is crucial for cultivating a supportive and productive work atmosphere. Collectively, these studies illuminate the significant role that emotions play in software development, influencing both the process and the results of collaborative projects.

\textbf{Datasets in ML-related projects.} The progress of sentiment analysis within software engineering has greatly benefited from the introduction and refinement of essential datasets. These datasets provide crucial benchmarks for understanding and enhancing the methodologies used in ML-related projects. Paullada et al.~\cite{paullada2021data} offer a comprehensive review of the role of datasets in machine learning research, emphasizing their critical importance and addressing the challenges associated with their collection and utilization. The authors advocate for more rigorous and careful dataset practices to improve research quality and applicability. Building on this foundation, Munaiah et al.~\cite{munaiah2017curating} introduced a pivotal dataset comprising 800 manually labeled software projects, evenly split between engineered and non-engineered categories. This dataset serves as a valuable resource for differentiating between types of software projects and forms the basis for further analytical work in the field. 

Expanding these resources, Pickerill et al.~\cite{pickerill2020phantom} contributed an additional 200 projects to the dataset, ensuring no overlap with the work of Munaiah et al.~\cite{munaiah2017curating}. This expansion enhances the depth and variety of data available for sentiment analysis within software engineering. Furthermore, Gonzalez et al.~\cite{gonzalez2020state} have identified ML-related projects that have garnered significant community engagement, enriching the understanding of how community interactions can influence software development practices. Additionally, Widysari et al.~\cite{widyasari2023niche} present the NICHE dataset, which includes 572 machine learning projects from GitHub. These projects are manually labeled to identify exemplary practices in software engineering. 
The focus on projects with at least 10,000 stars highlights those that are highly popular and influential within the ML community. 

\textbf{Code Quality Evaluation.} The evaluation of code quality is a critical aspect of software development, particularly in the context of ML projects where the integrity of the code can directly affect the performance and reliability of ML algorithms. Recent studies have emphasized the potential of sentiment analysis in enhancing code quality evaluation methods. Ronchieri et al.~\cite{ronchieri2019sentiment} have demonstrated the application of sentiment analysis in evaluating the quality of ML code repositories. By analyzing sentiment attributes extracted from Git commit messages, their research indicates that understanding the emotional content of these messages can aid in assessing overall code quality and even in evaluating the effectiveness of ML algorithms. 

Further exploring the tools and methodologies for code quality assessment, Daniel Guamán et al.~\cite{guaman2017sonarqube} investigated the use of the SQALE methodology combined with SonarQube for identifying technical debt through static analysis. Their study highlights the importance of adhering to established quality attributes and programming standards as defined by ISO 9126 and SQALE. SonarQube, in particular, is identified as a crucial tool for managing and controlling code quality across multiple dimensions, as detailed by J.L. Letouzey~\cite{letouzey2012sqale}. Moreover, the techniques employed for code analysis, such as clustering, static, and dynamic analysis, play a vital role in influencing various quality attributes, including maintainability, verifiability, and reusability. These methods are essential for understanding a system's architecture, as outlined by I. Pashov and M. Riebisch~\cite{1316725}. Static analysis, for instance, allows for the extraction of architectural models and traits from source code, providing insights into the system's framework, component dependencies, and potential coding issues, as discussed by Arias et al.~\cite{article}. Chong et al.~\cite{Chong} further detail how static analysis can facilitate the discovery of key software components and their interrelations. This includes the graphical representation of architectural elements and the metrics involved, such as class relation weights, static coupling, and cohesion, which are crucial for maintaining and enhancing code quality.

Despite these advancements, there remains a significant gap in directly correlating developers' sentiments with the quality of the code they produce. This gap suggests an unexplored area of research into how positive or negative sentiments among developers may impact the introduction of bugs, code maintainability, and overall software quality in ML projects. This study seeks to bridge this gap by investigating the complex interplay between developers' emotional dynamics and their influence on code quality practices within ML projects.
\section {Methodology}
\subsection{Data Collection}
We utilize the ML projects from the dataset curated by Widyasari et al.~\cite{widyasari2023niche}, which provides a comprehensive list of engineered ML projects developed using popular libraries such as Theano, PyTorch, and TensorFlow. This dataset includes vital project metadata like the number of stars, commit frequency, and the timestamps of the latest commits. To enhance the reliability of our analysis, we meticulously filter and manually label the projects, selecting 20 that met our high standards of engagement and relevance for sentiment analysis based on their star count and activity metrics.

We employ a crucial data collection process via the GitHub API~\footnote{https://docs.github.com/en/rest} to aggregate issue comments from selected repositories. The process involves specifying the target repository and utilizing the API endpoint to get issue comments. A personal access token is used to address potential rate limits. The script is configured to retrieve a predefined number of comments, adhering to the API’s limitation of 100 comments per page. During this process, comments are fetched and accumulated into a comprehensive list. This standardized procedure facilitates the extraction of a large dataset of issue comments, which can be analyzed to discern significant patterns of community interactions and contributions to the repository’s development.

\subsection{Data Pre-processing}

Practical data analysis requires good quality data that is easily interpretable. Preprocessing is a crucial step in achieving this, as it transforms raw data into a more refined form that can be easily analyzed. In the specific context of GitHub issue comments, the data may be noisy, making analysis difficult. Examples of noisy data include irrelevant or short comments that offer little analytical value, duplicate entries that can skew data distribution, and non-standard text elements such as emojis, URLs, and numbers that obscure meaningful patterns. 

The initial step involves the removal of short and duplicate comments to ensure data quality. Subsequently, we apply text normalization procedures, including converting emojis to text, eliminating URLs and numbers, and filtering out stop words and non-English terms. The remaining text is then processed to remove single characters and special symbols and convert all characters to lowercase. To further distill the data, we utilize the NLTK\footnote{https://www.nltk.org/} library’s stemming and lemmatization capabilities~\cite{hardeniya2016natural}, which reduce words to their root forms and authorize them. This meticulous preprocessing routine resulted in a cleansed and structured dataset primed for the subsequent stages of our analytical endeavor.

\subsection{Data Preparation}

Once the data has been thoroughly processed, we narrow down our focus to the top 20 repositories with a minimum of 5000 stars and 4500 comments. To facilitate deeper examination, we meticulously store the data for each of these repositories in separate CSV files. This procedure allows for a more comprehensive and insightful analysis later.

\subsection{Sentiment Analysis}
In the subsequent phase of our study, we conduct a sentiment analysis on the preprocessed dataset of GitHub issue comments to discern the developer’s emotional response to project developments. A diverse array of sentiment analysis tools was required to label the sentiment of comments within our dataset accurately. Recognizing the inherent variability in sentiment analysis outcomes, we employed five distinct NLP models. 

\begin{itemize}
    \item \textbf{VADER (Valence Aware Dictionary and Sentiment Reasoner)}: This model is adept at handling sentiments expressed in various domains such as social media texts. It calculates sentiment scores using a combination of a sentiment lexicon and grammatical rules~\cite{hutto2014vader}.
    \item \textbf{TextBlob}: This library provides a simple API for everyday NLP tasks, including sentiment analysis. It uses a lexicon of pre-labeled words to determine polarity and subjectivity scores~\cite{loria2018textblob}.
    \item \textbf{Pattern}: Another NLP library that, like TextBlob, uses a lexicon and returns polarity and subjectivity scores\footnote{https://github.com/clips/pattern}.
    \item \textbf{BERT (Bidirectional Encoder Representations from Transformers)}: A more advanced model that uses deep learning to understand the context of words in search queries. It is pre-trained on a large text corpus and fine-tuned for sentiment analysis~\cite{devlin2018bert}.
    \item \textbf{spaCy with SpacyTextBlob}: spaCy is a modern, reliable NLP framework, and SpacyTextBlob is an extension that adds sentiment analysis capabilities\footnote{https://spacy.io/universe/project/spacy-textblob}.
\end{itemize}

This ensemble approach allows us to harness the strengths of each model, from lexicon-based techniques to sophisticated transformer-based architectures. To reconcile the differing sentiment predictions and arrive at the most representative label for each comment, we implement a max voting mechanism. Max voting, a widely recognized decision-making method, selects the sentiment label that is most frequently predicted by the ensemble of models. This approach is encapsulated in the statistical function $mode$, which identifies the most common label among the predictions, defined as equations~\ref{eq:frequency_function} and \ref{eq:mode_definition}. When a tie occurs, indicating an equal frequency of sentiment labels, we resort to human evaluation to discern the most appropriate sentiment. This step ensures that the final sentiment label is determined with a nuanced understanding that may not be captured by automated methods alone. By leveraging max voting complemented by human judgment, we ensure that the final sentiment label reflects a robust consensus, thereby enhancing the reliability of our sentiment classification.

We define a function $(f)$ that counts the frequency of each sentiment label:

\begin{equation}\label{eq:frequency_function}
f(\text{label}) = \text{the number of times 'label' appears}
\end{equation}

The mode, $(M)$, is then the label for which $(f(label))$ is maximized:

\begin{equation}\label{eq:mode_definition}
M = \text{argmax}_{\text{label}} , f(\text{label})
\end{equation}

The $mode$ is the argument (in this case, the sentiment label) that maximizes the frequency function $(f)$. If there is more than one such argument, which means there is a tie, then the dataset is multimodal, and in this case human evaluation is used to decide the final label.

\subsection{Code Quality Analysis}
We employ SonarQube\footnote{https://docs.sonarsource.com/sonarqube/8.9} to analyze the code quality of the projects, leveraging its capability to classify issues by type and severity. SonarQube distinguishes five types of issues, each related to specific aspects of the code. Bugs, for instance, are code issues that could lead to erroneous behavior or crashes during execution, identified through pattern analysis within the codebase. Vulnerabilities denote code patterns susceptible to security breaches, with SonarQube scanning for risks such as SQL injection, cross-site scripting (XSS), and insecure data handling.

Code Smells pertain to maintainability issues, often indicative of technical debt. These do not directly cause bugs or vulnerabilities but complicate future maintenance and expansion of the codebase, with common examples including duplicated code, overly complex methods, and excessive parameters. Additionally, SonarQube pinpoints security hotspots, critical areas requiring review to prevent potential security issues.

Duplication issues are addressed by scanning for similar or identical blocks of code across the codebase, utilizing advanced algorithms capable of detecting duplication in extensive and varied files.

The severity of these issues is classified based on potential impacts on system performance or developer productivity. Blocker and critical issues pose significant threats to system stability, with blockers more likely to cause severe disruptions. Major issues, while less severe, substantially affect developer productivity, contrasting with minor issues that have minimal impact. Info issues, classified separately, represent neither bugs nor quality flaws.

Given the challenge in quantifying the severity of these issues, we focus on quantifying the percentage presence of bugs, vulnerabilities, code smells, and security hotspots within the analyzed projects.

\begin{figure*}[h!]
    \centering
    \includegraphics[width=0.9\textwidth]{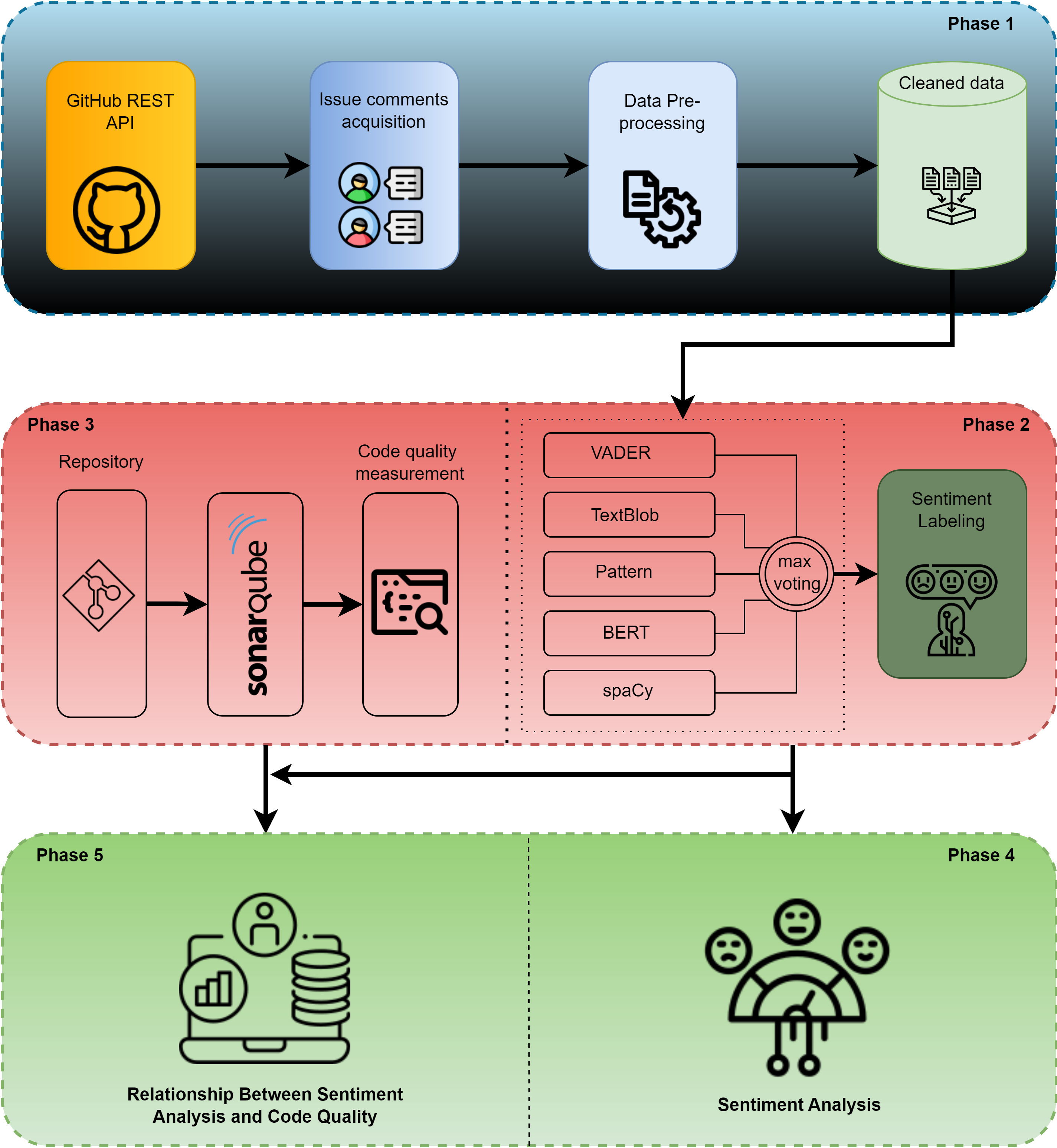} 
    \caption{The proposed framework presents a methodical diagram that illustrates the entire process, starting from data collection and then identifying sentiment analysis and the relationship between sentiment and code quality metrics.}
    \label{fig:system_architecure}
\end{figure*}

\subsection{System Architecture}

Our system architecture, depicted in Figure~\ref{fig:system_architecure}, is a well-structured and integrated framework that aims to analyze the correlation between community sentiment and code quality in software development. This process starts by extracting issue comments from GitHub repositories through the REST API. Then, we carry out a rigorous data preprocessing step to ensure the data is clean and well-structured. We employ a suite of NLP tools, which include VADER, TextBlob, Pattern, BERT, and spaCy, to label sentiments. These sentiments are then refined using a max voting approach to achieve consensus. Meanwhile, SonarQube evaluates the code quality of the repositories. Finally, the architecture integrates the sentiment analysis with the code quality metrics, providing a comprehensive understanding of how community feedback reflects the technical robustness of the software. This innovative approach offers a multi-dimensional perspective on the factors driving software excellence, enabling developers to develop more robust, high-quality software.

\subsection{Research Questions}
The aim of our research is to explore the intersection of developer sentiment and code quality within machine learning projects. The following research questions have been formulated to guide our investigation:

\begin{itemize}
    \item \textbf{RQ1:} What is the overall sentiment of developers involved in machine learning projects?
    \item \textbf{RQ2:} What do widely recognized machine learning projects measure up in terms of code quality?
    \item \textbf{RQ3:} How does developer sentiment correlate with bugs in the project?
    \item \textbf{RQ4:} What is the relationship between developer sentiment and the occurrence of code smells in ML projects?
    \item \textbf{RQ5:} How does developer sentiment affect the identification of security hotspots in the project?
    \item \textbf{RQ6:} Is there a correlation between developer sentiment and duplication issues in ML projects?
\end{itemize}
\section{Results}
In this section, we present the main findings of our research and answer the research questions we investigate.

\subsection{RQ1. What is the overall sentiment of developers involved in machine learning projects?}
\begin{figure*}[h!]
    \centering
    \includegraphics[width=1.0\textwidth]{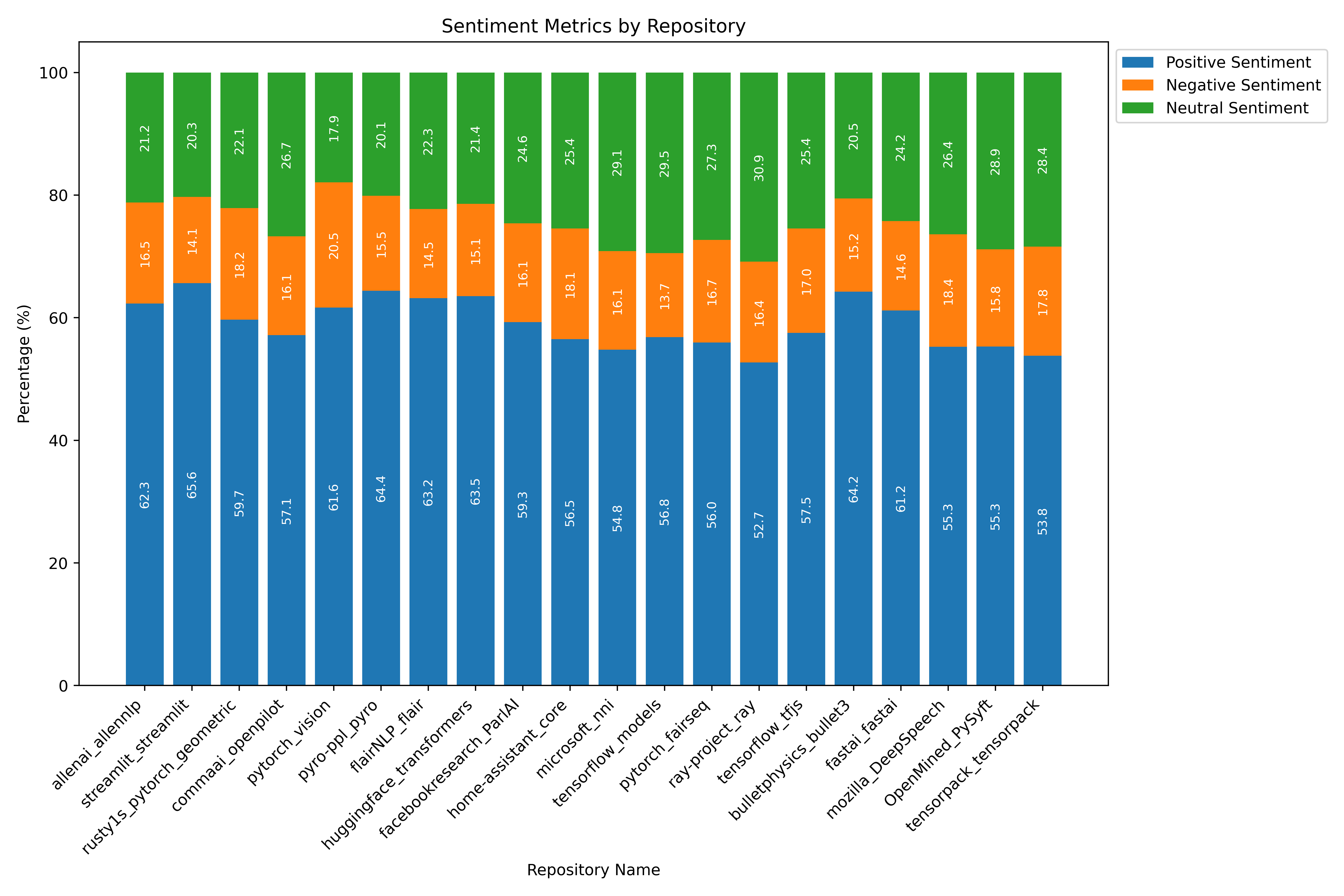} 
    \caption{Distribution of Sentiment Metrics by various machine learning projects.}
    \label{fig:sentiment_metrics}
\end{figure*}
Figure~\ref{fig:sentiment_metrics} shows that the overall sentiment expressed by developers in machine learning projects is predominantly positive. For instance, popular repositories like \textit{tensorflow} and \textit{pytorch} show that positive sentiments account for approximately 68\% and 64\%, respectively, suggesting that these projects provide a generally positive experience for developers. Conversely, negative sentiments average around 15\% across the projects, indicating a relatively lower occurrence. Neutral sentiments average about 17\%, showing that a significant portion of developers' sentiments remains neutral, neither particularly positive nor negative.

These results indicate that developers involved in machine learning projects tend to experience positive emotions, likely driven by the interesting challenges, a sense of achievement, and the positive aspects of teamwork within these projects. However, not all projects show a predominance of positive sentiments; some projects exhibit higher levels of negative or neutral sentiments, which could be influenced by factors such as project complexity, task difficulty, and team communication effectiveness.
Therefore, while the overall sentiment of developers in machine learning projects is generally positive, it's crucial to recognize that a spectrum of emotions exists, varying by project and circumstances. This recognition can provide valuable insights for project managers and team leaders in understanding and managing developers' emotions effectively. Recognizing the emotional state of developers can play an essential role in maintaining the health of the project and contributing to the developers' efficiency and satisfaction.

\subsection{RQ2. What do widely recognized machine learning projects measure up in terms of code quality in SonarQube?}
\begin{sidewaystable}[]
\centering
\caption{Code Quality Assessment of Machine Learning Projects}
\label{tab:code_quality}
\scriptsize
\setlength\tabcolsep{5pt}
\renewcommand{\arraystretch}{1.3}
\begin{tabular}{@{}p{4.0cm}p{1.2cm}cp{0.5cm}cp{0.5cm}cp{0.5cm}cp{0.5cm}c@{}}
\hline
\textbf{GitHub Repo} & \textbf{Lines of Code} & \textbf{Bugs} & \textbf{Bug Rating} & \textbf{Vulnerabilities} & \textbf{Vuln. Rating} & \textbf{Security Hotspots} & \textbf{Sec. Hotspots Rating} & \textbf{Code Smells} & \textbf{Code Smell Rating} & \textbf{Duplication} \\ \hline
allenai/allennlp & 58,830 & 197 & E & 0 & A & 92 & E & 630 & A & 2.70\% \\
streamlit/streamlit & 134,582 & 106 & E & 17 & E & 187 & E & 1,501 & A & 2.90\% \\
rusty1s/pytorch\_geometric & 102,452 & 120 & E & 2 & D & 59 & E & 815 & A & 5.30\% \\
commaai/openpilot & 131,299 & 50 & E & 6 & D & 162 & E & 8,056 & A & 0.50\% \\
pytorch/vision & 83,496 & 61 & E & 0 & A & 165 & E & 1,426 & A & 3.40\% \\
pyro-ppl/pyro & 88,486 & 89 & E & 0 & A & 8 & E & 1,554 & A & 5.90\% \\
flairNLP/flair & 32,751 & 19 & C & 0 & A & 56 & E & 598 & A & 3.20\% \\
huggingface/transformers & 890,921 & 342 & E & 6 & D & 267 & E & 11,098 & A & 30.60\% \\
facebookresearch/ParlAI & 157,108 & 128 & E & 1 & E & 385 & E & 3,065 & A & 1.80\% \\
home-assistant/core & 1,576,383 & 1,192 & E & 12 & E & 4,616 & E & 4,606 & A & 4.70\% \\
microsoft/nni & 123,547 & 147 & E & 0 & A & 225 & E & 2,373 & A & 7.40\% \\
tensorflow/models & 383,376 & 144 & E & 0 & A & 243 & E & 3,856 & A & 7.90\% \\
pytorch/fairseq & 146,075 & 0 & A & 96 & E & 173 & E & 1,868 & A & 7.30\% \\
ray-project/ray & 682,988 & 406 & E & 21 & E & 1,162 & E & 14,116 & A & 4.00\% \\
tensorflow/tfjs & 286,006 & 158 & D & 9 & E & 221 & E & 4,362 & A & 7.80\% \\
bulletphysics/bullet3 & 1,519,523 & 871 & E & 5 & D & 580 & E & 81,498 & A & 18\% \\
fastai/fastai & 14,159 & 26 & E & 0 & A & 1 & E & 556 & A & 0.10\% \\
mozilla/DeepSpeech & 220,524 & 34 & E & 0 & A & 139 & E & 33,088 & A & 66.40\% \\
OpenMined/PySyft & 65,489 & 74 & E & 17 & C & 196 & E & 1,041 & A & 1.80\% \\
tensorpack/tensorpack & 25,389 & 14 & E & 0 & A & 25 & E & 643 & A & 1.00\% \\ \hline
\end{tabular}
\end{sidewaystable}

The analysis based on SonarQube metrics provides a detailed view into the code quality of various notable machine learning projects, as summarized in Table~\ref{tab:code_quality}, which can be found at the end of the paper due to its size. This section discusses the implications of these metrics for overall project health and maintainability.

The metrics such as bugs, vulnerabilities, security hotspots, code smells, and duplication rates are examined to assess the quality of the code across different repositories.

\textit{\textbf{Bugs and Vulnerabilities:}} Variability in the number of bugs and vulnerabilities across projects indicates different levels of adherence to coding standards and testing rigor.

\textit{\textbf{Security Hotspots and Code Smells:}} High numbers in some projects suggest areas where security and maintenance practices could be improved.

\textit{\textbf{Duplication:}} Duplication rates vary, with higher percentages indicating potential areas for code consolidation and optimization.
The analysis indicates that while many machine learning projects are robust in functionality, there is a significant variation in code quality metrics that could impact long-term maintainability and scalability. Projects with high code smells and duplication rates might face challenges in future development phases unless addressed. On the other hand, projects with lower bugs and vulnerabilities are likely adhering well to coding standards and security practices.

\subsection{RQ3. How does developer sentiment correlate with bugs in the project?}
\begin{figure*}[h!]
    \centering
    \includegraphics[width=1.0\textwidth]{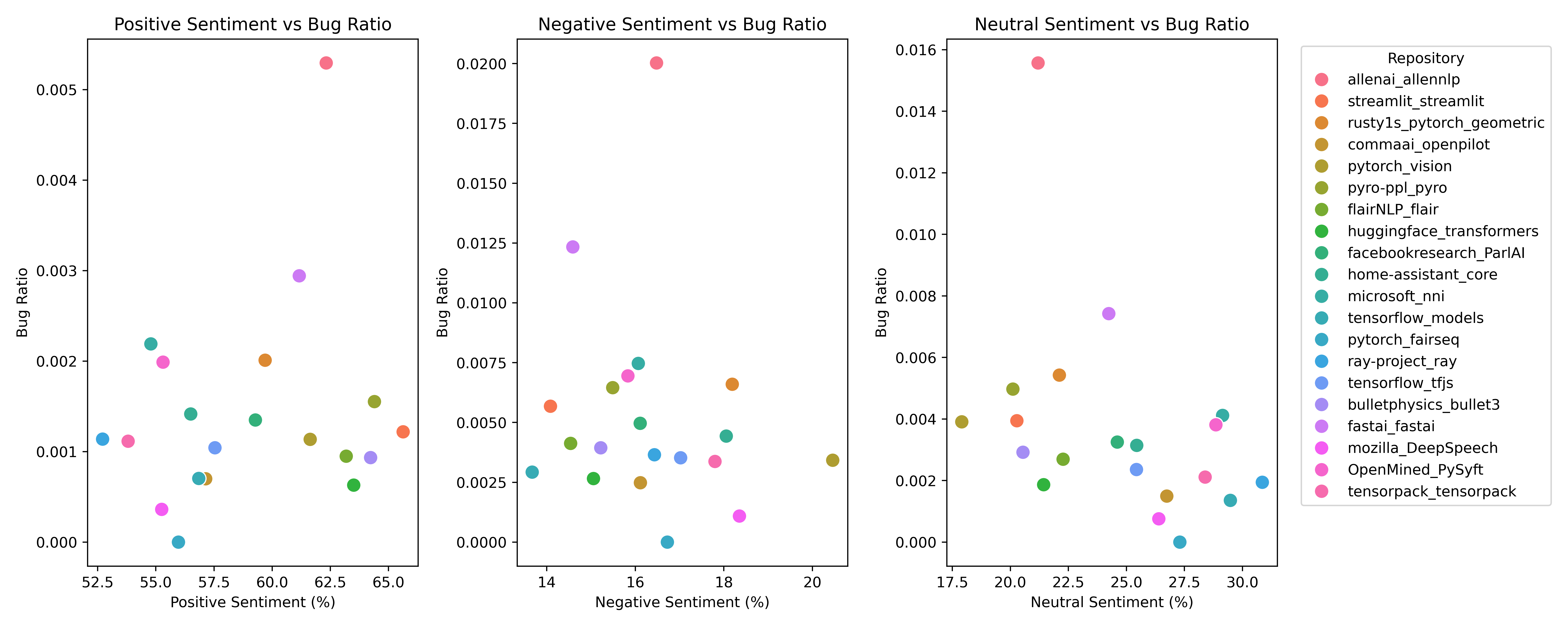} 
    \caption{Correlation between Developer Sentiments and Bug Ratios}
    \label{fig:bug_ratio_scatter}
\end{figure*}

Figure~\ref{fig:bug_ratio_scatter} illustrates the relationships between different types of sentiments—positive, negative, and neutral—expressed by developers and the bug ratio in various machine learning repositories. Analysis of these plots provides insights into how emotional states within development teams can influence software quality.

\textit{\textbf{Positive Sentiment vs. Bug Ratio:}} The left panel of Figure~\ref{fig:bug_ratio_scatter} illustrates lower bug ratios are observed with higher positive sentiments, suggesting that a positive emotional climate among developers correlates with fewer bugs. Projects like 'tensorflow\_models' and 'pytorch\_vision' exemplify this trend, indicating that motivated teams likely maintain higher code quality.

\textit{\textbf{Negative Sentiment vs. Bug Ratio:}} The middle panel of Figure~\ref{fig:bug_ratio_scatter} shows that higher bug ratios are associated with increased negative sentiments, indicating a possible link between developer dissatisfaction and a higher incidence of bugs. This trend may reflect the negative impact of stress or frustration on coding practices.

\textit{\textbf{Neutral Sentiment vs. Bug Ratio:}} The right panel of Figure~\ref{fig:bug_ratio_scatter} reveals that the relationship between neutral sentiments and bug ratios is less definitive, with a scattered data distribution. However, some clustering suggests that a moderate level of neutral sentiment often corresponds with an average bug ratio, possibly reflecting a balanced emotional state among developers.

\subsection{RQ4. What is the relationship between developer sentiment and the occurrence of code smells in ML projects?}
\begin{figure*}[h!]
    \centering
    \includegraphics[width=1.0\textwidth]{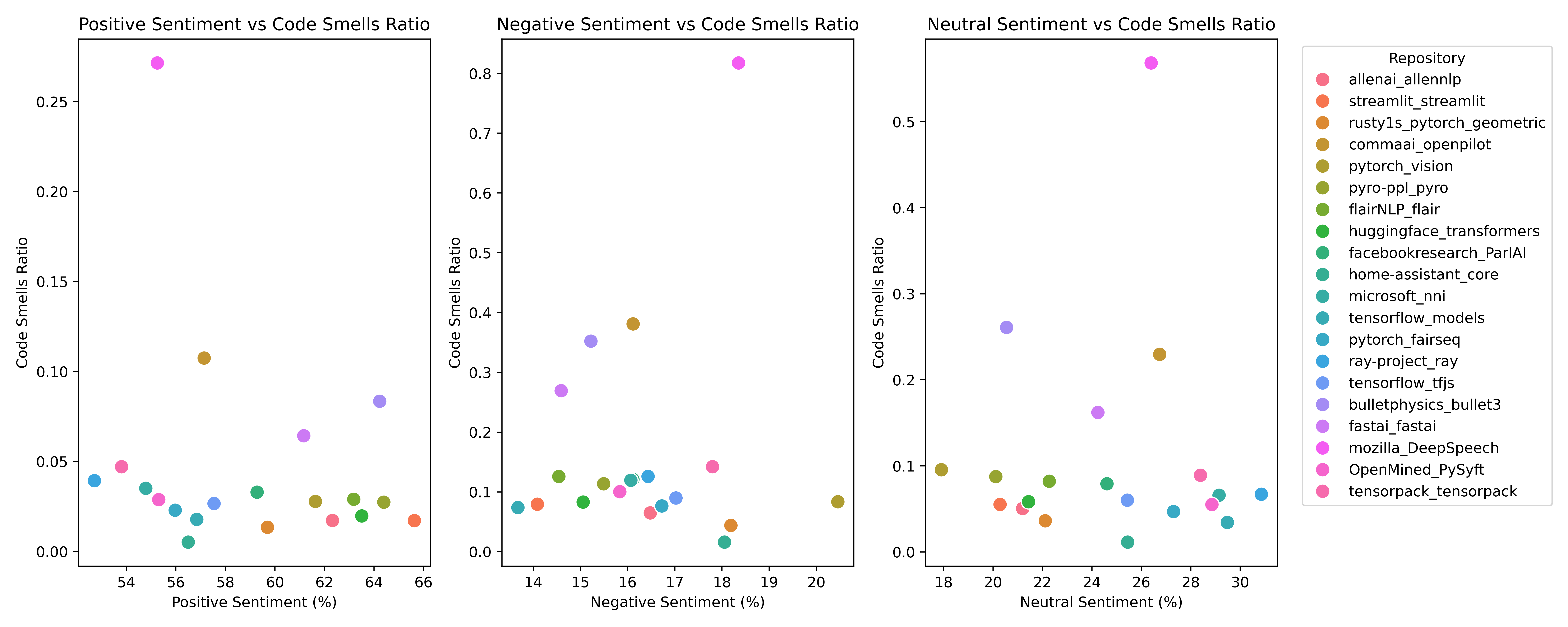}
    \caption{Correlation between Developer Sentiments and Code Smell Ratios}
    \label{fig:code_smells_ratio_scatter}
\end{figure*}

Figure~\ref{fig:code_smells_ratio_scatter} reveals how different types of developer sentiments—positive, negative, and neutral—affect the occurrence of code smells in various machine learning projects.

\textit{\textbf{Positive Sentiment vs. Code Smells Ratio:}} The left panel of Figure~\ref{fig:code_smells_ratio_scatter} indicates that higher positive sentiments correlate with lower code smells ratios, suggesting that a positive emotional climate within development teams can lead to higher code quality.

\textit{\textbf{Negative Sentiment vs. Code Smells Ratio:}} The middle panel of Figure~\ref{fig:code_smells_ratio_scatter} shows varied outcomes, with some points indicating higher code smells ratios associated with increased negative sentiments, suggesting that negative emotions might detract from coding standards.

\textit{\textbf{Neutral Sentiment vs. Code Smells Ratio:}} The right panel of Figure~\ref{fig:code_smells_ratio_scatter} presents the correlation is less definitive, indicating that neutral sentiments do not strongly correlate with code smells ratios, implying that other factors might influence code quality when emotional engagement is neutral.

From these observations, it is evident that developer sentiments have a significant impact on the occurrence of code smells in ML projects. Positive sentiments tend to be associated with fewer code smells, suggesting that maintaining a positive working environment can enhance code quality. On the other hand, negative sentiments are occasionally linked with higher occurrences of code smells, highlighting the need for project leaders to manage team dynamics carefully to mitigate negative impacts on code quality. Neutral sentiments show an ambiguous impact, suggesting that emotions are not the sole influencers of code quality and that other project-specific factors might play a critical role.

\subsection{RQ5. How does developer sentiment affect the identification of security hotspots in the project?}
\begin{figure*}[h!]
    \centering
    \includegraphics[width=1.0\textwidth]{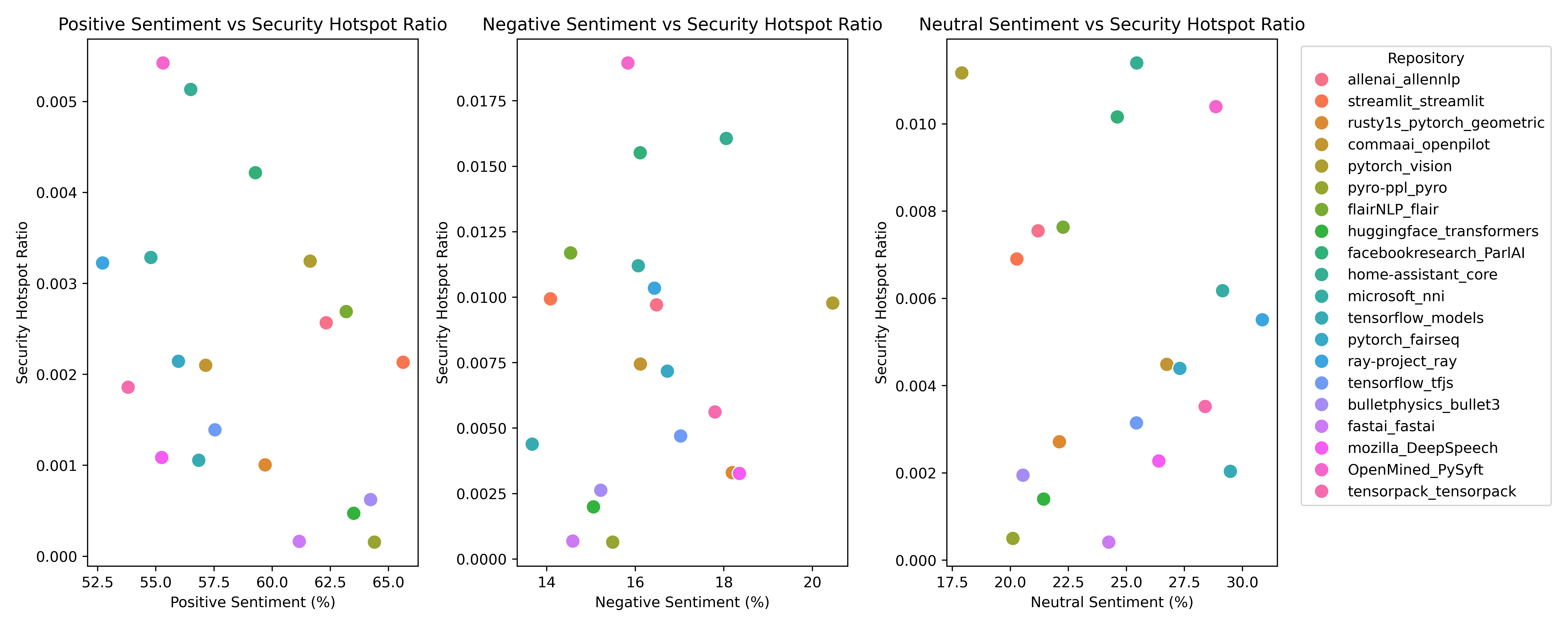}
    \caption{Correlation between Developer Sentiments and Security Hotspot Ratios}
    \label{fig:security_hotspot_ratio_scatter}
\end{figure*}

Figure~\ref{fig:security_hotspot_ratio_scatter} illustrates the correlation between different types of developer sentiments—positive, negative, and neutral—and the ratio of security hotspots in various machine learning projects. This analysis aids in understanding how the emotional climate within development teams might influence their security practices.

\textit{\textbf{Positive Sentiment vs. Security Hotspot Ratio:}} The left panel of Figure~\ref{fig:security_hotspot_ratio_scatter} shows a trend where higher positive sentiments correlate with lower security hotspot ratios, suggesting that positive environments may encourage better security practices. However, the correlation is not consistent across all data points.

\textit{\textbf{Negative Sentiment vs. Security Hotspot Ratio:}} The middle panel of Figure~\ref{fig:security_hotspot_ratio_scatter} presents higher negative sentiments often correlate with higher security hotspot ratios, indicating that negative emotional climates may lead to lapses in security practices.

\textit{\textbf{Neutral Sentiment vs. Security Hotspot Ratio:}} The right panel of Figure~\ref{fig:security_hotspot_ratio_scatter} illustrates varied outcomes with no clear trend, suggesting that neutral sentiments alone do not strongly predict the occurrence of security hotspots.

These observations imply that developer sentiments, particularly positive and negative, have implications for the identification of security hotspots in software projects. Positive sentiments might enhance security practices, potentially leading to fewer overlooked security risks. Conversely, negative sentiments could negatively impact security measures, possibly leading to more security hotspots. Neutral sentiments provide inconclusive results, indicating the need for further investigation into other influencing factors.

\subsection{RQ6. Is there a correlation between developer sentiment and duplication issues in ML projects?}
\begin{figure*}[h!]
    \centering
    \includegraphics[width=1.0\textwidth]{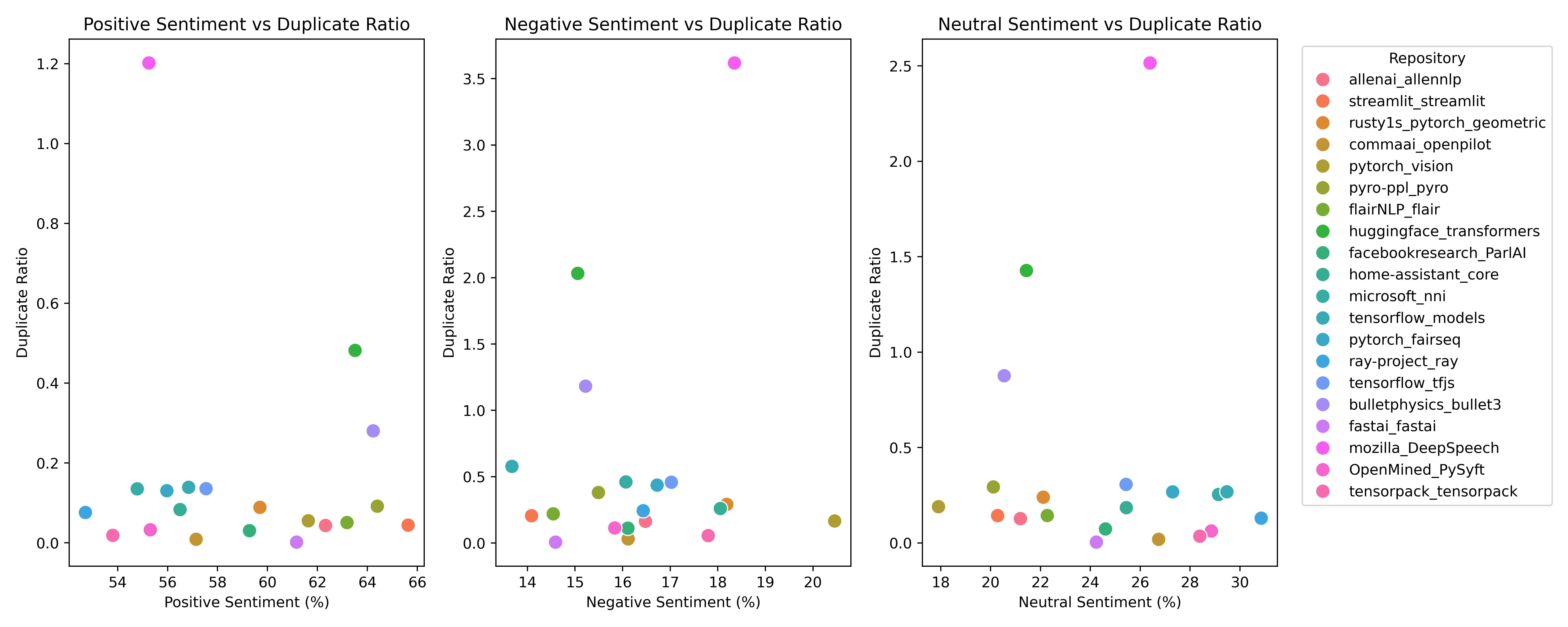}
    \caption{Correlation between Developer Sentiments and Duplication Ratios}
    \label{fig:duplicate_ratio_scatter}
\end{figure*}

Figure~\ref{fig:duplicate_ratio_scatter} reveals how different types of developer sentiments—positive, negative, and neutral—affect the duplication ratio in various machine learning projects.

\textit{\textbf{Positive Sentiment vs. Duplicate Ratio:}} The left panel of Figure~\ref{fig:duplicate_ratio_scatter} illustrates that higher positive sentiments generally correlate with lower duplication ratios, suggesting that a positive emotional climate may encourage more innovative and conscientious coding practices.

\textit{\textbf{Negative Sentiment vs. Duplicate Ratio:}} The middle panel of Figure~\ref{fig:duplicate_ratio_scatter} reveals that higher negative sentiments correlate with higher duplication ratios, indicating that negative emotions may lead to less careful coding practices, potentially increasing code redundancy.

\textit{\textbf{Neutral Sentiment vs. Duplicate Ratio:}} The right panel of Figure~\ref{fig:duplicate_ratio_scatter} shows a diverse range of outcomes, indicating that neutral emotions alone do not have a consistent impact on duplication issues, suggesting the influence of other project-specific factors.

These findings indicate that developer sentiments have a significant impact on code duplication in ML projects. Positive sentiments appear to reduce the likelihood of code duplication, potentially leading to more efficient and innovative coding practices. Negative sentiments, conversely, may contribute to an increase in code duplication, reflecting potentially less engagement or lower morale. Neutral sentiments do not show a consistent impact, suggesting that other factors may be more influential in these cases.

\section{Discussion}

Our study presents several findings regarding developer sentiment in ML projects and its correlation with various aspects of software quality. The overall sentiment of developers, based on contributions to popular repositories like TensorFlow and PyTorch, is predominantly positive. Positive sentiments make up around 64-68\% of the total, with only 15\% being negative. This suggests that most developers have a positive experience, possibly due to challenges, teamwork, and the nature of the work. However, some projects experience higher neutral or negative sentiments, which may be linked to complexities or communication issues.

Regarding code quality (measured through SonarQube metrics like bugs, vulnerabilities, and code smells), the results show considerable variability across ML projects. Projects with higher bugs and vulnerabilities might suffer from inconsistent coding standards, while higher security hotspots and code smells indicate areas needing improvement in security and maintenance practices. Duplication rates also vary, with some projects showing potential for code optimization. Generally, projects with fewer issues are likely adhering to better coding standards.

The study also explores the correlation between developer sentiment and project health. Positive sentiment tends to be associated with fewer bugs, lower code smells, and reduced duplication rates, reflecting higher code quality. Conversely, negative sentiment correlates with more bugs, code smells, and duplication issues, suggesting that dissatisfaction within teams can negatively impact software quality. Neutral sentiment shows less consistent patterns, indicating that other factors may play a role.

Lastly, developer sentiment also appears to influence the identification of security hotspots. Positive emotions may foster better security practices, while negative sentiments correlate with higher security risks. Neutral emotions do not provide conclusive trends, requiring further investigation into their role in project quality.
\section{Threats to Validity}

There are several potential threats to the validity of this research that should be considered:

\textbf{External Validity (Generalizability):} The dataset used in this study was curated from specific ML projects that primarily employ popular libraries such as Theano, PyTorch, and TensorFlow. As a result, the findings might not generalize to smaller, less popular ML projects or projects using other frameworks. Additionally, selecting only 20 repositories based on star count and activity metrics may introduce bias, as highly starred projects are more likely to have established communities and resources, which may not represent the experiences of developers working on less prominent projects.

\textbf{Internal Validity (Selection and Sampling Bias):} The selection of repositories for analysis was based on star count and engagement metrics, which might introduce selection bias. Repositories with more stars and higher activity could have better-organized teams or more frequent contributions, potentially skewing the sentiment and code quality assessments. Moreover, manually labeling projects, while increasing relevance, may introduce human bias during the selection process, which could affect the validity of the findings.

\textbf{Construct Validity (Sentiment and Code Quality Measures):} The use of multiple sentiment analysis tools, such as VADER, TextBlob, and BERT, could lead to inconsistent sentiment classifications due to the varying methodologies and lexicons used by these tools. Although the max voting mechanism aims to address this issue, it is possible that certain nuanced sentiments are misclassified, especially in the case of technical jargon commonly found in GitHub issue comments. Furthermore, the reliance on SonarQube for code quality assessment might not capture all relevant dimensions of code health, as certain qualitative aspects of code quality, such as developer productivity or collaboration efficiency, are difficult to quantify with automated tools.

Considering these threats helps contextualize the findings of this study and provides a foundation for future research to address these limitations, improving the robustness and generalizability of the results.
\section{Conclusion}
This research has underscored the significant impact of developer sentiments on code quality in machine learning projects. Our comprehensive analysis, incorporating advanced sentiment analysis techniques applied to diverse textual interactions within software development environments, reveals that positive sentiments among developers correlate strongly with higher code quality. Specifically, positive emotions within development teams are associated with fewer bugs and reduced instances of code smells. This suggests that a supportive and encouraging environment can directly enhance technical outcomes, fostering not only productivity but also a higher standard of code craftsmanship.

Conversely, we observed that negative sentiments often correlate with an increase in code issues, such as duplication and security vulnerabilities. This indicates the potential adverse effects of a negative emotional climate on the technical integrity and security of software projects. These findings highlight the necessity for project managers and team leaders to cultivate positive emotional environments to mitigate risks and enhance project outcomes.

Furthermore, our study contributes to a deeper understanding of the role of emotional intelligence in software engineering. By demonstrating that the emotional states of developers can significantly influence the technical quality of their output, our research supports the integration of emotional intelligence training and practices in software engineering teams, especially in contexts that demand high levels of innovation and collaboration, such as machine learning projects.

In conclusion, while technical skills and coding expertise are paramount, the emotional dimensions of software development teams also play a critical role in the success of projects. Future research should explore strategic interventions aimed at enhancing emotional intelligence across development teams, potentially including personalized training modules, team-building exercises, and the incorporation of tools that monitor and manage emotional health to promote a more holistic approach to improving software quality.

\bibliographystyle{plain}
\bibliography{references}


\begin{thebibliography}{27}
\ifx \bisbn   \undefined \def \bisbn  #1{ISBN #1}\fi
\ifx \binits  \undefined \def \binits#1{#1}\fi
\ifx \bauthor  \undefined \def \bauthor#1{#1}\fi
\ifx \batitle  \undefined \def \batitle#1{#1}\fi
\ifx \bjtitle  \undefined \def \bjtitle#1{#1}\fi
\ifx \bvolume  \undefined \def \bvolume#1{\textbf{#1}}\fi
\ifx \byear  \undefined \def \byear#1{#1}\fi
\ifx \bissue  \undefined \def \bissue#1{#1}\fi
\ifx \bfpage  \undefined \def \bfpage#1{#1}\fi
\ifx \blpage  \undefined \def \blpage #1{#1}\fi
\ifx \burl  \undefined \def \burl#1{\textsf{#1}}\fi
\ifx \doiurl  \undefined \def \doiurl#1{\url{https://doi.org/#1}}\fi
\ifx \betal  \undefined \def \betal{\textit{et al.}}\fi
\ifx \binstitute  \undefined \def \binstitute#1{#1}\fi
\ifx \binstitutionaled  \undefined \def \binstitutionaled#1{#1}\fi
\ifx \bctitle  \undefined \def \bctitle#1{#1}\fi
\ifx \beditor  \undefined \def \beditor#1{#1}\fi
\ifx \bpublisher  \undefined \def \bpublisher#1{#1}\fi
\ifx \bbtitle  \undefined \def \bbtitle#1{#1}\fi
\ifx \bedition  \undefined \def \bedition#1{#1}\fi
\ifx \bseriesno  \undefined \def \bseriesno#1{#1}\fi
\ifx \blocation  \undefined \def \blocation#1{#1}\fi
\ifx \bsertitle  \undefined \def \bsertitle#1{#1}\fi
\ifx \bsnm \undefined \def \bsnm#1{#1}\fi
\ifx \bsuffix \undefined \def \bsuffix#1{#1}\fi
\ifx \bparticle \undefined \def \bparticle#1{#1}\fi
\ifx \barticle \undefined \def \barticle#1{#1}\fi
\bibcommenthead
\ifx \bconfdate \undefined \def \bconfdate #1{#1}\fi
\ifx \botherref \undefined \def \botherref #1{#1}\fi
\ifx \url \undefined \def \url#1{\textsf{#1}}\fi
\ifx \bchapter \undefined \def \bchapter#1{#1}\fi
\ifx \bbook \undefined \def \bbook#1{#1}\fi
\ifx \bcomment \undefined \def \bcomment#1{#1}\fi
\ifx \oauthor \undefined \def \oauthor#1{#1}\fi
\ifx \citeauthoryear \undefined \def \citeauthoryear#1{#1}\fi
\ifx \endbibitem  \undefined \def \endbibitem {}\fi
\ifx \bconflocation  \undefined \def \bconflocation#1{#1}\fi
\ifx \arxivurl  \undefined \def \arxivurl#1{\textsf{#1}}\fi
\csname PreBibitemsHook\endcsname

\bibitem[\protect\citeauthoryear{Taboada}{2016}]{taboada2016sentiment}
\begin{barticle}
\bauthor{\bsnm{Taboada}, \binits{M.}}:
\batitle{Sentiment analysis: An overview from linguistics}.
\bjtitle{Annual Review of Linguistics}
\bvolume{2},
\bfpage{325}--\blpage{347}
(\byear{2016})
\end{barticle}
\endbibitem

\bibitem[\protect\citeauthoryear{Ahmed et~al.}{2021}]{ahmed2021detecting}
\begin{barticle}
\bauthor{\bsnm{Ahmed}, \binits{M.S.}},
\bauthor{\bsnm{Aurpa}, \binits{T.T.}},
\bauthor{\bsnm{Anwar}, \binits{M.M.}}:
\batitle{Detecting sentiment dynamics and clusters of twitter users for
  trending topics in covid-19 pandemic}.
\bjtitle{Plos one}
\bvolume{16}(\bissue{8}),
\bfpage{0253300}
(\byear{2021})
\end{barticle}
\endbibitem

\bibitem[\protect\citeauthoryear{Lin et~al.}{2018}]{lin2018sentiment}
\begin{bchapter}
\bauthor{\bsnm{Lin}, \binits{B.}},
\bauthor{\bsnm{Zampetti}, \binits{F.}},
\bauthor{\bsnm{Bavota}, \binits{G.}},
\bauthor{\bsnm{Di~Penta}, \binits{M.}},
\bauthor{\bsnm{Lanza}, \binits{M.}},
\bauthor{\bsnm{Oliveto}, \binits{R.}}:
\bctitle{Sentiment analysis for software engineering: How far can we go?}
In: \bbtitle{Proceedings of the 40th International Conference on Software
  Engineering},
pp. \bfpage{94}--\blpage{104}
(\byear{2018})
\end{bchapter}
\endbibitem

\bibitem[\protect\citeauthoryear{Asghar et~al.}{2017}]{asghar2017lexicon}
\begin{barticle}
\bauthor{\bsnm{Asghar}, \binits{M.Z.}},
\bauthor{\bsnm{Khan}, \binits{A.}},
\bauthor{\bsnm{Ahmad}, \binits{S.}},
\bauthor{\bsnm{Qasim}, \binits{M.}},
\bauthor{\bsnm{Khan}, \binits{I.A.}}:
\batitle{Lexicon-enhanced sentiment analysis framework using rule-based
  classification scheme}.
\bjtitle{PloS one}
\bvolume{12}(\bissue{2}),
\bfpage{0171649}
(\byear{2017})
\end{barticle}
\endbibitem

\bibitem[\protect\citeauthoryear{Hasan et~al.}{2018}]{hasan2018machine}
\begin{barticle}
\bauthor{\bsnm{Hasan}, \binits{A.}},
\bauthor{\bsnm{Moin}, \binits{S.}},
\bauthor{\bsnm{Karim}, \binits{A.}},
\bauthor{\bsnm{Shamshirband}, \binits{S.}}:
\batitle{Machine learning-based sentiment analysis for twitter accounts}.
\bjtitle{Mathematical and computational applications}
\bvolume{23}(\bissue{1}),
\bfpage{11}
(\byear{2018})
\end{barticle}
\endbibitem

\bibitem[\protect\citeauthoryear{Le}{2019}]{le2019hybrid}
\begin{bchapter}
\bauthor{\bsnm{Le}, \binits{T.}}:
\bctitle{A hybrid method for text-based sentiment analysis}.
In: \bbtitle{2019 International Conference on Computational Science and
  Computational Intelligence (CSCI)},
pp. \bfpage{1392}--\blpage{1397}
(\byear{2019}).
\bcomment{IEEE}
\end{bchapter}
\endbibitem

\bibitem[\protect\citeauthoryear{Keuning et~al.}{2023}]{keuning2023systematic}
\begin{bchapter}
\bauthor{\bsnm{Keuning}, \binits{H.}},
\bauthor{\bsnm{Jeuring}, \binits{J.}},
\bauthor{\bsnm{Heeren}, \binits{B.}}:
\bctitle{A systematic mapping study of code quality in education}.
In: \bbtitle{Proceedings of the 2023 Conference on Innovation and Technology in
  Computer Science Education V. 1},
pp. \bfpage{5}--\blpage{11}
(\byear{2023})
\end{bchapter}
\endbibitem

\bibitem[\protect\citeauthoryear{Biswas et~al.}{2020}]{biswas2020achieving}
\begin{bchapter}
\bauthor{\bsnm{Biswas}, \binits{E.}},
\bauthor{\bsnm{Karabulut}, \binits{M.E.}},
\bauthor{\bsnm{Pollock}, \binits{L.}},
\bauthor{\bsnm{Vijay-Shanker}, \binits{K.}}:
\bctitle{Achieving reliable sentiment analysis in the software engineering
  domain using bert}.
In: \bbtitle{2020 IEEE International Conference on Software Maintenance and
  Evolution (ICSME)},
pp. \bfpage{162}--\blpage{173}
(\byear{2020}).
\bcomment{IEEE}
\end{bchapter}
\endbibitem

\bibitem[\protect\citeauthoryear{Gachechiladze et~al.}{2017}]{Gachechiladze}
\begin{bchapter}
\bauthor{\bsnm{Gachechiladze}, \binits{D.}},
\bauthor{\bsnm{Lanubile}, \binits{F.}},
\bauthor{\bsnm{Novielli}, \binits{N.}},
\bauthor{\bsnm{Serebrenik}, \binits{A.}}:
\bctitle{Anger and its direction in collaborative software development},
pp. \bfpage{11}--\blpage{14}
(\byear{2017}).
\doiurl{10.1109/ICSE-NIER.2017.18}
\end{bchapter}
\endbibitem

\bibitem[\protect\citeauthoryear{Guzman and Bruegge}{2013}]{Guzman}
\begin{bchapter}
\bauthor{\bsnm{Guzman}, \binits{E.}},
\bauthor{\bsnm{Bruegge}, \binits{B.}}:
\bctitle{Towards emotional awareness in software development teams},
pp. \bfpage{671}--\blpage{674}
(\byear{2013}).
\doiurl{10.1145/2491411.2494578}
\end{bchapter}
\endbibitem

\bibitem[\protect\citeauthoryear{Cao and Park}{2017}]{Cao2017UnderstandingGE}
\begin{bchapter}
\bauthor{\bsnm{Cao}, \binits{L.}},
\bauthor{\bsnm{Park}, \binits{E.H.}}:
\bctitle{Understanding goal-directed emotions in agile software development
  teams}.
In: \bbtitle{Americas Conference on Information Systems}
(\byear{2017}).
\burl{https://api.semanticscholar.org/CorpusID:12282379}
\end{bchapter}
\endbibitem

\bibitem[\protect\citeauthoryear{Fountaine and Sharif}{2017}]{Fountaine}
\begin{bchapter}
\bauthor{\bsnm{Fountaine}, \binits{A.}},
\bauthor{\bsnm{Sharif}, \binits{B.}}:
\bctitle{Emotional awareness in software development: Theory and measurement}.
In: \bbtitle{2017 IEEE/ACM 2nd International Workshop on Emotion Awareness in
  Software Engineering (SEmotion)},
pp. \bfpage{28}--\blpage{31}
(\byear{2017}).
\doiurl{10.1109/SEmotion.2017.12}
\end{bchapter}
\endbibitem

\bibitem[\protect\citeauthoryear{Paullada et~al.}{2021}]{paullada2021data}
\begin{botherref}
\oauthor{\bsnm{Paullada}, \binits{A.}},
\oauthor{\bsnm{Raji}, \binits{I.D.}},
\oauthor{\bsnm{Bender}, \binits{E.M.}},
\oauthor{\bsnm{Denton}, \binits{E.}},
\oauthor{\bsnm{Hanna}, \binits{A.}}:
Data and its (dis) contents: A survey of dataset development and use in machine
  learning research.
Patterns
\textbf{2}(11)
(2021)
\end{botherref}
\endbibitem

\bibitem[\protect\citeauthoryear{Munaiah et~al.}{2017}]{munaiah2017curating}
\begin{barticle}
\bauthor{\bsnm{Munaiah}, \binits{N.}},
\bauthor{\bsnm{Kroh}, \binits{S.}},
\bauthor{\bsnm{Cabrey}, \binits{C.}},
\bauthor{\bsnm{Nagappan}, \binits{M.}}:
\batitle{Curating github for engineered software projects}.
\bjtitle{Empirical Software Engineering}
\bvolume{22},
\bfpage{3219}--\blpage{3253}
(\byear{2017})
\end{barticle}
\endbibitem

\bibitem[\protect\citeauthoryear{Pickerill et~al.}{2020}]{pickerill2020phantom}
\begin{barticle}
\bauthor{\bsnm{Pickerill}, \binits{P.}},
\bauthor{\bsnm{Jungen}, \binits{H.J.}},
\bauthor{\bsnm{Ochodek}, \binits{M.}},
\bauthor{\bsnm{Ma{\'c}kowiak}, \binits{M.}},
\bauthor{\bsnm{Staron}, \binits{M.}}:
\batitle{Phantom: Curating github for engineered software projects using
  time-series clustering}.
\bjtitle{Empirical Software Engineering}
\bvolume{25},
\bfpage{2897}--\blpage{2929}
(\byear{2020})
\end{barticle}
\endbibitem

\bibitem[\protect\citeauthoryear{Gonzalez et~al.}{2020}]{gonzalez2020state}
\begin{bchapter}
\bauthor{\bsnm{Gonzalez}, \binits{D.}},
\bauthor{\bsnm{Zimmermann}, \binits{T.}},
\bauthor{\bsnm{Nagappan}, \binits{N.}}:
\bctitle{The state of the ml-universe: 10 years of artificial intelligence \&
  machine learning software development on github}.
In: \bbtitle{Proceedings of the 17th International Conference on Mining
  Software Repositories},
pp. \bfpage{431}--\blpage{442}
(\byear{2020})
\end{bchapter}
\endbibitem

\bibitem[\protect\citeauthoryear{Widyasari et~al.}{2023}]{widyasari2023niche}
\begin{bchapter}
\bauthor{\bsnm{Widyasari}, \binits{R.}},
\bauthor{\bsnm{Yang}, \binits{Z.}},
\bauthor{\bsnm{Thung}, \binits{F.}},
\bauthor{\bsnm{Sim}, \binits{S.Q.}},
\bauthor{\bsnm{Wee}, \binits{F.}},
\bauthor{\bsnm{Lok}, \binits{C.}},
\bauthor{\bsnm{Phan}, \binits{J.}},
\bauthor{\bsnm{Qi}, \binits{H.}},
\bauthor{\bsnm{Tan}, \binits{C.}},
\bauthor{\bsnm{Tay}, \binits{Q.}}, \betal:
\bctitle{Niche: A curated dataset of engineered machine learning projects in
  python}.
In: \bbtitle{2023 IEEE/ACM 20th International Conference on Mining Software
  Repositories (MSR)},
pp. \bfpage{62}--\blpage{66}
(\byear{2023}).
\bcomment{IEEE}
\end{bchapter}
\endbibitem

\bibitem[\protect\citeauthoryear{Ronchieri
  et~al.}{2019}]{ronchieri2019sentiment}
\begin{bchapter}
\bauthor{\bsnm{Ronchieri}, \binits{E.}},
\bauthor{\bsnm{Juric}, \binits{R.}},
\bauthor{\bsnm{Canaparo}, \binits{M.}}:
\bctitle{Sentiment analysis for software code assessment}.
In: \bbtitle{2019 IEEE Nuclear Science Symposium and Medical Imaging Conference
  (NSS/MIC)},
pp. \bfpage{1}--\blpage{2}
(\byear{2019}).
\bcomment{IEEE}
\end{bchapter}
\endbibitem

\bibitem[\protect\citeauthoryear{Guaman}{}]{guaman2017sonarqube}
\begin{botherref}
\oauthor{\bsnm{Guaman}, \binits{D.}}:
Sonarqube as a tool to identify software metrics and technical debt in the
  source code through static analysis
\end{botherref}
\endbibitem

\bibitem[\protect\citeauthoryear{Letouzey}{2012}]{letouzey2012sqale}
\begin{bchapter}
\bauthor{\bsnm{Letouzey}, \binits{J.-L.}}:
\bctitle{The sqale method for evaluating technical debt}.
In: \bbtitle{2012 Third International Workshop on Managing Technical Debt
  (MTD)},
pp. \bfpage{31}--\blpage{36}
(\byear{2012}).
\bcomment{IEEE}
\end{bchapter}
\endbibitem

\bibitem[\protect\citeauthoryear{Pashov and Riebisch}{2004}]{1316725}
\begin{bchapter}
\bauthor{\bsnm{Pashov}, \binits{I.}},
\bauthor{\bsnm{Riebisch}, \binits{M.}}:
\bctitle{Using feature modeling for program comprehension and software
  architecture recovery}.
In: \bbtitle{Proceedings. 11th IEEE International Conference and Workshop on
  the Engineering of Computer-Based Systems, 2004.},
pp. \bfpage{406}--\blpage{417}
(\byear{2004}).
\doiurl{10.1109/ECBS.2004.1316725}
\end{bchapter}
\endbibitem

\bibitem[\protect\citeauthoryear{Arias et~al.}{2011}]{article}
\begin{barticle}
\bauthor{\bsnm{Arias}, \binits{T.}},
\bauthor{\bsnm{Avgeriou}, \binits{P.}},
\bauthor{\bsnm{America}, \binits{P.}},
\bauthor{\bsnm{Blom}, \binits{K.}},
\bauthor{\bsnm{Bachynskyy}, \binits{S.}}:
\batitle{A top-down strategy to reverse architecting execution views for a
  large and complex software-intensive system: An experience report}.
\bjtitle{Sci. Comput. Program.}
\bvolume{76},
\bfpage{1098}--\blpage{1112}
(\byear{2011})
\doiurl{10.1016/j.scico.2010.11.008}
\end{barticle}
\endbibitem

\bibitem[\protect\citeauthoryear{Chong et~al.}{2013}]{Chong}
\begin{barticle}
\bauthor{\bsnm{Chong}, \binits{C.Y.}},
\bauthor{\bsnm{Lee}, \binits{S.}},
\bauthor{\bsnm{Ling}, \binits{T.C.}}:
\batitle{Efficient software clustering technique using an adaptive and
  preventive dendrogram cutting approach}.
\bjtitle{Information and Software Technology}
\bvolume{55},
\bfpage{1994}--\blpage{2012}
(\byear{2013})
\doiurl{10.1016/j.infsof.2013.07.002}
\end{barticle}
\endbibitem

\bibitem[\protect\citeauthoryear{Hardeniya et~al.}{2016}]{hardeniya2016natural}
\begin{bbook}
\bauthor{\bsnm{Hardeniya}, \binits{N.}},
\bauthor{\bsnm{Perkins}, \binits{J.}},
\bauthor{\bsnm{Chopra}, \binits{D.}},
\bauthor{\bsnm{Joshi}, \binits{N.}},
\bauthor{\bsnm{Mathur}, \binits{I.}}:
\bbtitle{Natural Language Processing: Python and NLTK}.
\bpublisher{Packt Publishing Ltd}, \blocation{???}
(\byear{2016})
\end{bbook}
\endbibitem

\bibitem[\protect\citeauthoryear{Hutto and Gilbert}{2014}]{hutto2014vader}
\begin{bchapter}
\bauthor{\bsnm{Hutto}, \binits{C.}},
\bauthor{\bsnm{Gilbert}, \binits{E.}}:
\bctitle{Vader: A parsimonious rule-based model for sentiment analysis of
  social media text}.
In: \bbtitle{Proceedings of the International AAAI Conference on Web and Social
  Media},
vol. \bseriesno{8},
pp. \bfpage{216}--\blpage{225}
(\byear{2014})
\end{bchapter}
\endbibitem

\bibitem[\protect\citeauthoryear{Loria et~al.}{2018}]{loria2018textblob}
\begin{barticle}
\bauthor{\bsnm{Loria}, \binits{S.}}, \betal:
\batitle{textblob documentation}.
\bjtitle{Release 0.15}
\bvolume{2}(\bissue{8}),
\bfpage{269}
(\byear{2018})
\end{barticle}
\endbibitem

\bibitem[\protect\citeauthoryear{Devlin et~al.}{2018}]{devlin2018bert}
\begin{botherref}
\oauthor{\bsnm{Devlin}, \binits{J.}},
\oauthor{\bsnm{Chang}, \binits{M.-W.}},
\oauthor{\bsnm{Lee}, \binits{K.}},
\oauthor{\bsnm{Toutanova}, \binits{K.}}:
Bert: Pre-training of deep bidirectional transformers for language
  understanding.
arXiv preprint arXiv:1810.04805
(2018)
\end{botherref}
\endbibitem

\end{thebibliography}

\end{document}